\documentclass[12pt]{article}
\usepackage{graphicx}

\begin{document}

\begin{center}

{\Large \bf ABCD Matrices as Similarity Transformations of
Wigner Matrices and Periodic Systems in Optics }

\vspace{7mm}

S. Ba{\c s}kal \footnote{electronic address:
baskal@newton.physics.metu.edu.tr}\\
Department of Physics, Middle East Technical University,
06531 Ankara, Turkey

\vspace{5mm}

Y. S. Kim \footnote{electronic address: yskim@physics.umd.edu}\\
Department of Physics, University of Maryland,\\
College Park, Maryland 20742

\end{center}

\begin{abstract}
The beam transfer matrix, often called the $ABCD$ matrix, is a
two-by-two matrix with unit determinant, and with three independent
parameters.  It is noted that this matrix cannot always be
diagonalized.  It can however be brought by rotation to a matrix
with equal diagonal elements.  This equi-diagonal matrix can then
be squeeze-transformed to a rotation, to a squeeze, or to one of the
two shear matrices.  It is noted that these one-parameter matrices
constitute the basic elements of the Wigner's little group for space-time
symmetries of elementary particles.  Thus every $ABCD$ matrix can
be written as a similarity transformation of one of the Wigner
matrices, while the transformation matrix is a rotation preceded
by a squeeze.  This mathematical property enables us to compute
scattering processes in periodic systems.  Laser cavities and
multilayer optics are discussed in detail.  For both cases, it
is shown possible to write the one-cycle transfer matrix as a
similarity transformation of one of the Wigner matrices.  It is
thus possible to calculate the $ABCD$ matrix for an arbitrary
number of cycles.
\end{abstract}
\

\newpage

\section{Introduction}\label{intro}
Without too much exaggeration, modern physics consists of harmonic
oscillators and two-by-two matrices.  Ray optics is mostly the
physics of two-by-two matrices.  The beam transfer matrix, usually
called the $ABCD$ matrix, is a two-by-two matrix with unit determinant
and with three independent parameters.

In mathematics, the group of matrices with this property is called
the two-dimensional symplectic group or $Sp(2)$.  Indeed, the $Sp(2)$
group has been playing a pivotal role both in classical and quantum
optics.  It is the basic language for lens optics~\cite{bk01,bk03},
multilayer optics~\cite{monzon00,gk03}, laser cavities~\cite{bk02}, as
well as squeezed states of light in quantum optics~\cite{yuen76,knp91}.

This group is locally isomorphic to the group of Lorentz transformations
applicable to one time dimension and two space dimensions.  This Lorentz
group consists of three-by-three matrices.  Since the algebra of
two-by-two matrices is much simpler than those of three-by-three, the
$Sp(2)$ group provides a very convenient computational tool for Lorentz
transformations in physics~\cite{kn83ajp}.

While our understanding of the Pauli matrices and the rotation group
is thorough, we are not yet completely familiar with the matrices of the
$Sp(2)$ group~\cite{guill84,lang85}.  Do we know how to diagonalize the
$ABCD$ matrix?  We shall point out in this paper that, unlike the case
of the rotation matrices, it is not always possible to digonalize the $ABCD$
matrix by a similarity transformation.  We shall show however that every
$ABCD$ matrix can be rotated to a matrix with equal diagonal elements.

We shall then show that this equi-diagonal matrix can be squeezed to
one of the four one-parameter matrices consisting rotation, squeeze,
and two shear matrices.  They take the form
\begin{eqnarray}\label{101}
&{}& \pmatrix{\cos\theta^* & -\sin\theta^* \cr
              \sin\theta^* & \cos\theta^*} ,   \quad
\pmatrix{\cosh\lambda^* & \sinh\lambda^* \cr
                \sinh\lambda^* & \cosh\lambda^*} , \nonumber\\[1.0ex]
&{}& \pmatrix{1 & -\gamma^* \cr 0 & 1}, \quad
                \pmatrix{1 & 0 \cr \gamma^* & 1} .
\end{eqnarray}
One of the two triangular matrices in the above expression can be
obtained from the other by rotation, but it is more convenient to
use both of them.

It should be noted that these four matrices form the four basic
matrices of Wigner's little group for space-time symmetries for
elementary particles~\cite{wig39,knp86}.  We shall therefore call
them Wigner matrices, and use $W$ collectively for these four
matrices.

These matrices have equal diagonal elements, but not all of them
are diagonalizable.  The first matrix is a rotation matrix, whose
properties are well known.  However, its eigenvalues are complex,
and therefore it cannot be diagonalized if all the matrices in the
group are to be real.  The second matrix is symmetric
and can be diagonalized by a rotation.  The third and fourth
matrices are triangular and cannot be diagonalized.

On the other hand, these matrices share one convenient feature
with the diagonal matrix.  Namely, for the first matrix,
\begin{eqnarray}\label{102}
&{}& \pmatrix{\cos\theta_1 & -\sin\theta_1 \cr
   \sin\theta_1 & \cos\theta_1}
 \pmatrix{\cos\theta_2 & -\sin\theta_2
 \cr \sin\theta_2 & \cos\theta_2}  \nonumber \\[1.0ex]
&{}& \hspace{10mm} = \pmatrix{\cos\left(\theta_1 + \theta_2\right) &
 -\sin\left(\theta_1 + \theta_2\right) \cr
 \sin\left(\theta_1 + \theta_2\right)
           & \cos\left(\theta_1 + \theta_2\right)} ,
\end{eqnarray}
and thus
\begin{equation}\label{103}
\pmatrix{\cos\theta & -\sin\theta \cr \sin\theta & \cos\theta}^N =
\pmatrix{\cos(N\theta) & -\sin(N\theta) \cr \sin(N\theta)
                & \cos(N\theta)} ,
\end{equation}
The multiplication of two matrices results in the addition of
parameters.  This was called ``slide-rule'' property in
reference~\cite{gk03}, but it is more appropriate to call it
``logarithmic property.''  The remaining three matrices in
equation~(\ref{101}) have this logarithmic property.

We shall show in this paper that the $ABCD$ matrix can be brought
to one of the Wigner matrices by a similarity transformation, and
that the similarity transformation is a rotation followed by a
squeeze.

This mathematical result can be useful in calculating the $ABCD$
matrix for periodic systems.  In this paper, we study laser
cavities and multilayer optics in detail.  In both cases, we reduce
the multi-cycles system into one cycle.  Both subjects have been
extensively discussed in the literature.

Recently, the present authors used the method of Wigner's little
group to calculate the $ABCD$ matrix for laser cavities~\cite{bk01}.
However, their cycle had to start from the midway between the two
mirrors.  In this paper, our cycle will start at an arbitrary
point.

In 2003, Georgieva and Kim used a method based on the Lorentz group
to study multilayer optics~\cite{gk03}.  These authors also end
up with the inconvenience of starting their cycle from the midway
in one of the layers.  We eliminate this inconvenience.

In Sec.~\ref{eigen}, the eigenvalues of the $ABCD$ matrix is
discussed in detail.  They can be real, complex, and one.  Not
all of them can be brought to a diagonal form, especially if we
restrict the matrices to be real.  It is noted however that
every $ABCD$ matrix can be brought to a form with equal diagonal
elements by a rotation.
In Sec.~\ref{simil}, we construct a similarity transformation which
will bring the $ABCD$ matrix into one of the four Wigner matrices.
It is shown that this matrix is a rotation matrix followed by a
squeeze matrix.  It is also shown that these four different
Wigner matrices can be combined into one analytical expression.

The present formalism is applicable to one-dimensional periodic
systems.  In Sec.~\ref{cav}, we study laser cavities in detail.
In Sec.~\ref{multi}, we show how useful Wigner matrices are in
computing the scattering matrix for multilayer optics.

\section{Eigenvalues of the ABCD matrix}\label{eigen}
Let us start with the two-by-two matrix
\begin{equation}\label{abcd00}
\pmatrix{A  & B \cr C  & D} ,
\end{equation}
where the elements $A, B, C, D$ are real.  The determinant
of this matrix is one and thus $(AD - B C) = 1.$

The eigenvalue equation for this two-by-two matrix
is only a quadratic equation, and it is always soluble.
The eigenvalue equation becomes
\begin{equation}\label{abcd01}
E^2 - (A + D)E + AD - BC  = 0 ,
\end{equation}
and the eigenvalues are
\begin{equation}\label{abcd02}
E_{\pm} = \frac{1}{2} \left(A + D
   \pm \sqrt{(A - D)^2 + 4BC} \right) .
\end{equation}
If the quantity inside the square root sign is negative and
$(A + D)/2$ is smaller than one, then
the eigenvalues become
\begin{equation}\label{abcd003}
E_{\pm} = \exp{\left(\pm i\theta^*\right)},
\end{equation}
with
\begin{equation}\label{abcd04}
\cos\theta^* = \frac{A + D}{2} .
\end{equation}
If the quantity inside the square root sign is positive, and
$(A + D)/2$ is greater than one.  The eigenvalues are
\begin{equation}
E_{\pm} = \exp{\left(\pm \lambda^*\right)},
\end{equation}
with
\begin{equation}\label{abcd05}
\cosh\lambda^* = \frac{A + D}{2}.
\end{equation}
If the quantity inside the square root sign vanishes,
\begin{equation}\label{abcd06}
E_{\pm} = 1.
\end{equation}
These are all well known, and the purpose of this section is not
to repeat these trivial procedures.

Although we are used to diagonal matrices with eigenvalues as
diagonal elements, it is very cumbersome to deal with the above three separate
cases.  In order to find a common ground, we propose to use the
matrix of the form with equal diagonal elements:
\begin{equation}\label{abcd07}
 \pmatrix{J & F \cr G & J} ,
\end{equation}
and propose to bring the $ABCD$ matrix of equation~(\ref{abcd00})
to this form by a similarity transformation.  Since the trace is
preserved under similarity transformations,
\begin{equation}\label{abcd08}
J = \frac{A + D}{2} .
\end{equation}
Since the determinant of the $ABCD$ matrix is one,
\begin{equation}\label{abcd09}
J^2 - FG = 1 .
\end{equation}

In this paper, we started with the $ABCD$ matrix with three
independent parameters.  Because of the condition on the
diagonal elements, the $JFGJ$ matrix has two independent
parameters.  If the off diagonal elements $F$ and $G$ have
opposite signs, it can be written as
\begin{equation}\label{abcd10}
  \pmatrix{\cos\theta^*  & -e^\eta \sin\theta^* \cr
  e^{-\eta} \sin\theta^* &  \cos\theta^*} .
\end{equation}
If $F$ and $G$ have the same sign, it can take the form
\begin{equation}\label{abcd11}
  \pmatrix{\cosh\lambda^*  & e^\eta \sinh\lambda^* \cr
  e^{-\eta} \sinh\lambda^* &  \cos\lambda^*} .
\end{equation}
If $F$ or $G$ vanishes, we can write the $JFGJ$ matrix as
\begin{equation}\label{abcd12}
\pmatrix{1 & e^\eta \gamma^* \cr 0 & 1}, \quad or \quad
\pmatrix{1 & 0 \cr e^{-\eta} \gamma^* & 1} .
\end{equation}

Let us go back to equation~(\ref{101}), the above matrices are
similarity transformations of the Wigner matrices with
the transformation matrix
\begin{equation}\label{abcd13}
\pmatrix{e^{\eta/2} & 0 \cr 0 & e^{-\eta/2}} ,
\end{equation}
which expands one axis and contracts the other in the
two-dimensional space.  This is a squeeze transformation.
In general, symmetric two-by-two matrices perform squeeze
transformations.  Indeed, the matrices in equations~(\ref{abcd10})
to (\ref{abcd12}) are ``squeezed'' Wigner matrices.

\section{Similarity Transformations}\label{simil}
In order to find a similarity transformation which will bring
the $ABCD$ matrix of equation~(\ref{abcd00}) to the equi-diagonal
matrix of equation~(\ref{abcd07}), let us try a rotation
\begin{equation}\label{sim00}
 \pmatrix{\cos(\delta/2) & \sin(\delta/2)  \cr
   -\sin(\delta/2) & \cos(\delta/2) }
  \pmatrix{A & B \cr C & D}
  \pmatrix{\cos(\delta/2) & -\sin(\delta/2)  \cr
  \sin(\delta/2) & \cos(\delta/2) } .
\end{equation}
We can calculate the angle $\delta$ first.  The result is
\begin{equation}\label{sim01}
    \tan\delta = \frac{D - A}{B + C}  .
\end{equation}
Using this angle, we can then compute $F$ and $G$.  They are
\begin{eqnarray}\label{sim02}
&{}& F = \frac{1}{2}\left(B - C + \sqrt{(A - D)^2 + (B + C)^2}\right) ,
                   \nonumber \\[1ex]
&{}& G = \frac{1}{2}\left(C - B + \sqrt{(A - D)^2 + (B + C)^2}\right) .
\end{eqnarray}

If $J$ is smaller than one, we can write $J = \cos\theta^*,$ and use
equation~(\ref{abcd10}) for the $JFGJ$ matrix, with
\begin{eqnarray}\label{sim04}
&{}& \cos\theta^* = \frac{A + D}{2}, \nonumber \\[1ex]
&{}& e^{2\eta} = \frac{C - B - \sqrt{(A - D)^2 + (B + C)^2}}
             {C - B + \sqrt{(A - D)^2 + (B + C)^2}} ,
\end{eqnarray}
and the Wigner matrix should take the form
\begin{equation}\label{sim04a}
\pmatrix{\cos\theta^* & -\sin\theta^* \cr
  \sin\theta^* & \cos\theta^*} ,
\end{equation}
which is one of the four matrices given in equation~(\ref{101}).

If $J$ is greater than 1, we should use equation~(\ref{abcd11}),
with
\begin{eqnarray}\label{sim06}
&{}& \cosh\lambda^* = \frac{A + D}{2} , \nonumber \\[1ex]
&{}& e^{2\eta} = \frac{B - C + \sqrt{(A - D)^2 + (B + C)^2}}
             {C - B + \sqrt{(A - D)^2 + (B + C)^2}} .
\end{eqnarray}
The Wigner matrix in this case is
\begin{equation}
\pmatrix{\cosh\lambda^* & \sin\lambda^*  \cr
 \cosh\lambda^* & \sinh\lambda^*},
\end{equation}
which is also given in equation~(\ref{101}).

If $J = 1$, the diagonal elements are one, and the off-diagonal
element $F$ or $G$ has to vanish.  As a consequence, the $ABCD$
matrix becomes
\begin{equation}\label{sim07}
\pmatrix{1 & B - C \cr 0 & 1} \quad or \quad
\pmatrix{1 & 0 \cr C - B & 1} ,
\end{equation}
and
\begin{equation}\label{sim08}
C - B = e^{\eta}\gamma^*, \quad or \quad
C - B = e^{-\eta}\gamma^*,
\end{equation}
which are similarity transformations of the triangularr matrices
given in equation~(\ref{101}).

Therefore, $ABCD$ is a similarity transformation of one of the
Wigner matrices given in equation~(\ref{101}).  We thus write
\begin{equation}\label{sim09}
ABCD = S W S^{-1} ,
\end{equation}
where $W$ is the Wigner matrix given in equation~(\ref{101}), and
the similarity transformation matrix is
\begin{equation}\label{sim10}
  S = \pmatrix{\cos(\delta/2) & -\sin(\delta/2) \cr
       \sin(\delta/2) & \cos(\delta/2)}
      \pmatrix{e^{\eta/2} & 0 \cr 0 & e^{-\eta/2} } .
\end{equation}
This is a rotation matrix preceded by a squeeze matrix.

If there are four different Wigner matrices, calculations become
very cumbersome in practical applications.  There are no problems
if only one Wigner matrix, such as that of equation~(\ref{sim04a}),
is applicable throughout the problem as in the case of laser
cavities discussed in Sec.~\ref{cav}.

If two or more forms are involved in one problem, as in the case
of multilayer optics discussed in Sec.~\ref{multi}, the problem
could be uncontrollable.  Thus, we are led to consider one
mathematical expression for all four Wigner matrices.
For this purpose, let us consider the form
\begin{equation}\label{bgm01}
  \pmatrix{\cos(\theta/2) & -\sin(\theta/2) \cr
             \sin(\theta/2) & \cos(\theta/2)}
\pmatrix{\cosh\lambda & \sinh\lambda \cr \sinh\lambda & \cosh\lambda}
\pmatrix{\cos(\theta/2) & -\sin(\theta/2) \cr
             \sin(\theta/2) & \cos(\theta/2)} .
\end{equation}
This is a special case of the Bargmann decomposition~\cite{barg47},
and its physical interpretation in special relativity is discussed
in the literature~\cite{hk88}.  In this paper, we consider this
form purely for mathematical convenience for discussing periodic
systems.  After multiplication, the above expression can be
compressed into
\begin{equation}\label{bgm02}
\pmatrix{(\cosh\lambda)\cos\theta &
            -(\cosh\lambda)\sin\theta + \sinh\lambda \cr
    (\cosh\lambda)\sin\theta + \sinh\lambda
           & (\cosh\lambda)\cos\theta} ,
\end{equation}
with equal diagonal elements.  This matrix has two independent
parameters.
Indeed, this is another form of the $JFGJ$ matrix with
\begin{eqnarray}\label{bgm03}
&{}& J = (\cosh\lambda)\cos\theta ,   \nonumber\\[1ex]
&{}& F = -(\cosh\lambda)\sin\theta + \sinh\lambda , \nonumber \\[1ex]
&{}& G = (\cosh\lambda)\sin\theta + \sinh\lambda .
\end{eqnarray}
When $\lambda = 0,$ it becomes
\begin{equation}\label{bgm04}
\pmatrix{\cos\theta & -\sin\theta \cr \sin\theta & \cos\theta} ,
\end{equation}
while it takes the form
\begin{equation}\label{bgm05}
\pmatrix{\cosh\lambda & \sinh\lambda \cr \sinh\lambda & \cosh\lambda} ,
\end{equation}
when $\theta = 0.$

When $\tanh\lambda = \pm\sin\theta,$ the equi-diagonal matrix  of
equation~(\ref{bgm02}) becomes
\begin{equation}
\pmatrix{1 & 0 \cr 2\sinh\eta & 1}, \qquad
\pmatrix{1 & -2\sinh\eta \cr 0 & 1},
\end{equation}
respectively.

\section{Laser Cavities}\label{cav}

The laser cavity consists of two concave mirrors separated by distance
$s$.  The mirror matrix takes the form
\begin{equation}\label{las01}
\pmatrix{1 & 0 \cr -2/R & 1} ,
\end{equation}
where $R$ is the radius of the concave mirror.  The separation matrix
is
\begin{equation}\label{las02}
\pmatrix{1 & s \cr 0 & 1} .
\end{equation}
If we start the cycle from one of the two mirrors
one complete cycle consists of
\begin{equation}\label{las03}
\pmatrix{1 & 0 \cr -2/R & 1} \pmatrix{1 & s \cr 0 & 1}
\pmatrix{1 & 0 \cr -2/R & 1} \pmatrix{1 & s \cr 0 & 1} .
\end{equation}
If we start the at the position $d$ from the mirror, then the
complete cycle becomes
\begin{equation}\label{las03a}
\pmatrix{1 & d \cr 0 & 1}
\pmatrix{1 & 0 \cr -2/R & 1} \pmatrix{1 & s \cr 0 & 1}
\pmatrix{1 & 0 \cr -2/R & 1} \pmatrix{1 & s - d \cr 0 & 1} ,
\end{equation}
since
\begin{equation}
\pmatrix{1 & s \cr 0 & 1} = \pmatrix{1 & s - d \cr 0 & 1}
 \pmatrix{1 & d \cr 0 & 1} .
\end{equation}
Thus, one complete cycle consists of two repeated applications of
the half cycle, which can be written as
\begin{equation}\label{las04}
L = \pmatrix{1 & d \cr 0 & 1} \pmatrix{1 & 0 \cr -2/R & 1}
   \pmatrix{1 & s - d \cr 0 & 1},
\end{equation}
which becomes
\begin{equation}\label{las05}
L = \pmatrix{1 - 2d/R  & (1 - 2d/R)(s - d) + d \cr
 - 2/R & 1 - 2(s - d)/R} .
\end{equation}

One complete cycle then becomes $L^2.$  There is however inconvenient
feature of this expression is that the off-diagonal elements have
different dimensions, while the diagonal elements are dimensionless.
In order to deal with this problem, we write this expression as
a similarity transformation
\begin{equation}\label{las06}
 \pmatrix{\sqrt{s} & 0 \cr 0 & 1/\sqrt{s}}
 \pmatrix{1 - 2d/R & (1 - 2d/R)(s - d)/s + d/s \cr
 - 2s/R & 1 - 2(s - d)/R}
   \pmatrix{1/\sqrt{s} & 0 \cr 0 & \sqrt{s}} .
\end{equation}
The elements of the middle matrix are now dimensionless.  Let us now
use the notations $a = d/s$ and $b = 2/R$.  Then we can work with
the normalized $L$, which takes the form
\begin{equation}\label{las07}
L = \pmatrix{ 1 - ab &  1 -ab(1 - a) \cr -b & 1 - b(1 - a)} ,
\end{equation}
where $s = 1$, according to the normalization defined in
equation~(\ref{las06}).  We can now bring this form into a equi-diagonal
matrix.  We can write this expression in the
form of equation~(\ref{sim00}) with
\begin{eqnarray}\label{las08}
&{}& \tan\delta = \frac{b - 2ab}{1 - b - ab(1 - a)} , \nonumber \\[1ex]
&{}& F = \frac{1}{2}\left((1 + b) - ab(1 - a)
   + \sqrt{(b - 2ab)^2 + \left[1 - b\left(1 + a -
   a^2\right)\right]^2}\right), \hspace{7mm}   \nonumber \\[1ex]
&{}& G = \frac{1}{2}\left(ab(1 - a) - (1 + b) + \sqrt{(b - 2ab)^2
    + \left[1 - b\left(1 + a - a^2\right)\right]^2}\right) .
\end{eqnarray}
We can now write the $ABCD$ matrix for one complete cycle as a similarity
transformation
\begin{equation}
   \left[S R\left(2\theta^* \right) S^{-1}\right]^2 =
   S R\left(4\theta^* \right) S^{-1}  ,
\end{equation}
and
\begin{equation}
R\left(2\theta^*\right) = \pmatrix{\cos\theta^*  & -\sin\theta^* \cr
                         \sin\theta^*  & \cos\theta^* },
\end{equation}
with
\begin{equation}
\cos\theta^* = 1 - \frac{b}{2} .
\end{equation}
The similarity transformation matrix $S$ is in equation~(\ref{sim10}) and
takes the form
\begin{equation}
S = \pmatrix{e^{\eta/2}\cos(\delta/2) & -e^{-\eta/2}\sin(\delta/2) \cr
     e^{\eta/2}\sin(\delta/2) & e^{-\eta/2}\cos(\delta/2)} ,
\end{equation}
with
\begin{equation}
e^{2\eta} = -\frac{F}{G} ,
\end{equation}
where $F, G$, and $\delta$ are given in equation~(\ref{las08}).

Thus, for $N$ cycles, the $ABCD$ matrix becomes
\begin{equation}
   \left[S R\left(2\theta^* \right) S^{-1}\right]^{2N} =
   S R\left(4N\theta^* \right) S^{-1}  ,
\end{equation}
where we have used the logarithmic property of the Wigner matrix.

When the cycle starts from the midpoint in the cavity, $a = 1/2$,
and the half-cycle matrix becomes
\begin{equation}\label{las09}
L = \pmatrix{1 - b/2 &  1 - b/4 \cr -b & 1 - b/2} ,
\end{equation}
and $\delta$ becomes zero. We do not need the rotation matrix,
and this matrix can be written as
\begin{equation}\label{las10}
L = \pmatrix{\cos\theta^* & e^{\eta}\sin\theta^* \cr
   -e^{-\eta}\sin\theta^* &   \cos\theta^*} ,
\end{equation}
with
$$
\cos\theta^* = 1 - \frac{b}{2}, \qquad
                            e^{2\eta} = \frac{4 - b}{4b} .
$$
The matrix $L$ can of course be written as a similarity transformation
\begin{equation}\label{las11}
L = \pmatrix{e^{\eta/2}& 0 \cr 0 & e^{-\eta/2}}
\pmatrix{\cos\theta^* & \sin\theta^* \cr -\sin\theta^* & \cos\theta^*}
   \pmatrix{e^{-\eta/2}& 0 \cr 0 & e^{\eta/2}} .
\end{equation}
For the laser consisting of $N$ cycles, the $ABCD$ matrix becomes
\begin{equation}\label{las12}
L^{2N} = \pmatrix{e^{\eta/2}& 0 \cr 0 & -e^{-\eta/2}}
\pmatrix{\cos\left(N\theta^*\right) & \sin\left(N\theta^*\right)
     \cr -\sin\left(N\theta^*\right) & \cos\left(N\theta^*\right)}
   \pmatrix{e^{-\eta/2}& 0 \cr 0 & -e^{\eta/2}} ,
\end{equation}
and thus
\begin{equation}\label{las13}
L^{2N} = \pmatrix{\cos\left(N\theta^*\right) &
             e^{\eta}\sin\left(N\theta^*\right) \cr
 -e^{-\eta}\sin\left(N\theta^*\right) & \cos\left(N\theta^*\right)} .
\end{equation}
This is the result we obtained in our previous paper on laser
cavities~\cite{bk02}.  The point of this paper is we can start the
cycle from an arbitrary point, by introducing the parameter $\delta.$

When the cycle begins from one of the lenses, $a = 0$, and the $L$ matrix
becomes
\begin{equation}\label{las14}
L = \pmatrix{ 1   &  1  \cr -b & 1 - b} .
\end{equation}
This expression can first be brought to the equi-diagonal
form by a rotation matrix with the rotation angle
\begin{equation}
\tan \delta = \left(\frac{b}{1 - b}\right) ,
\end{equation}
according to equation~(\ref{sim01}).  The equi-diagonal matrix then
takes the form of equation~(\ref{las10}) or equation~(\ref{las11})
with
\begin{eqnarray}
&{}& \cos\theta^* = 1 - \frac{b}{2} ,  \nonumber \\[1.0ex]
&{}&  e^{2\eta} = \frac{(1 + b) + \sqrt{b^2 +(1 - b)^2}}
                 {(1 + b) - \sqrt{b^2 +(1 - b)^2}} .
\end{eqnarray}
Indeed the $L$ matrix of equation~(\ref{las14}) is a similarity
transformation of the Wigner matrix
\begin{equation}
\pmatrix{\cos\theta^* & \sin\theta^* \cr -\sin\theta^* & \cos\theta^*} ,
\end{equation}
with the transformation matrix of the form given in
equation~(\ref{sim10}), which is a rotation preceded by a squeeze.
Then it is straight-forward to calculate the $ABCD$ matrix for $N$
cycles.

\section{Multilayer Optics}\label{multi}
In multilayer optics, we are led to consider two beams moving in
opposite directions~\cite{azzam77}.  One is the incident beam
and the other is the reflected beam.  We can represent them as
a two component column matrix
\begin{equation}
\pmatrix{ E_{+}e^{ikx} \cr E_{-} e^{-ikx}} ,
\end{equation}
where the upper and lower components correspond to the incoming
and reflected beams respectively.
For a given frequency, the wave number depends on the index of the
refraction.  Thus, if the beam travels along the distance d, the
column matrix should be multiplied by the two-by-two
matrix~\cite{azzam77}
\begin{equation}
\pmatrix{e^{i\alpha/2} & 0 \cr 0 & e^{-i\alpha/2}} ,
\end{equation}
where $\alpha/2 = kd $.  Thus, the propagation matrices for
two different media can be represented by
\begin{equation}
\pmatrix{e^{i\alpha_1/2} & 0 \cr  0  & e^{-i\alpha_1/2} }, \qquad
     \pmatrix{e^{i\alpha_2/2} & 0 \cr 0  & e^{-i\alpha_2/2}} ,
\end{equation}
respectively, with $\alpha_1/2 = k_1d$ and $\alpha_2/2 = k_2d.$

If the beam propagates along the first medium and meets the boundary
at the second medium, it will be partially reflected and partially
transmitted.  The boundary matrix is~\cite{azzam77,monzon00}
\begin{equation}
\pmatrix{\cosh(\mu/2)  &  \sinh(\mu/2)  \cr
                \sinh(\mu/2)  &  \cosh(\mu/2) } ,
\end{equation}
with
\begin{equation}
\cosh(\mu/2) = 1/t_{12}, \qquad \sinh(\mu/2) = r_{12}/t_{12} ,
\end{equation}
where $t_{12}$ and $r_{12}$ are the transmission and reflection
coefficients respectively, and they satisfy
$\left(r_{12}^2 + t_{12}^2\right) = 1.$
The boundary matrix for the second to first medium is the
inverse of the above matrix and can be written as
\begin{equation}
    \pmatrix{\cosh(\mu/2)  &  -\sinh(\mu/2)  \cr
                -\sinh(\mu/2)  &  \cosh(\mu/2) } .
\end{equation}
Thus one complete cycle, starting from the second medium,
consists of
\begin{eqnarray}
&{}& \pmatrix{\cosh(\mu/2)  &  \sinh(\mu/2)  \cr
                \sinh(\mu/2)  &  \cosh(\mu/2) }
    \pmatrix{e^{i\alpha_1/2} & 0 \cr  0  & e^{-i\alpha_1/2} }
                                \nonumber \\[1ex]
&{}& \hspace{10mm} \times
   \pmatrix{\cosh(\mu/2)  &  -\sinh(\mu/2)  \cr
        - \sinh(\mu/2)  &  \cosh(\mu/2)}
\pmatrix{e^{i\alpha_2/2} & 0 \cr 0  & e^{-i\alpha_2/2}} .
\end{eqnarray}
This expression can be unitarily transformed to~\cite{gk01}
\begin{eqnarray}\label{m55}
&{}& M = \pmatrix{e^{\mu/2}  & 0 \cr 0 & e^{-\mu/2}}
    \pmatrix{\cos(\alpha_1/2)  & -\sin(\alpha_1/2) \cr
            \sin(\alpha_1/2)  & \cos(\alpha_1/2)}
  \pmatrix{e^{-\mu/2}  & 0 \cr 0 & e^{\mu/2}}   \nonumber \\[1ex]
&{}& \hspace{10mm} \times \pmatrix{\cos(\alpha_2/2)
     & -\sin(\alpha_2/2) \cr \sin(\alpha_2/2) & \cos(\alpha_2/2)} .
\end{eqnarray}
The first three matrices in the above expression can be compressed
to
\begin{equation}\label{m66}
\pmatrix{\cos(\alpha_1/2)  &  - e^\mu\sin(\alpha_1/2) \cr
           e^{-\mu} \sin(\alpha_1/2)  & \cos(\alpha_1/2)} ,
\end{equation}
and thus, according to equation~(\ref{bgm02}) and
equation~(\ref{bgm04}), to
\begin{equation}
  \pmatrix{\cos(\theta_1) \cosh\lambda &
     -\sin(\theta_1)\cosh\lambda + \sinh\lambda \cr
 \sin(\theta_1) \cosh\lambda + \sinh\lambda &
                  \cos(\theta_1) \cosh\lambda} ,
\end{equation}
with
\begin{eqnarray}
&{}& \cosh\lambda = (\cosh\mu)\sqrt{1 - \cos^2(\alpha_1/2)\tanh^2\mu} ,
                                  \nonumber \\[1ex]
&{}& \cos\theta_1 = \frac{\cos(\alpha_1/2)}
                {(\cosh\mu)\sqrt{1 - \cos^2(\alpha_1/2)\tanh^2\mu}}.
\end{eqnarray}
Then the matrix $M$ of equation~(\ref{m55}) can be written as
\begin{eqnarray}
&{}& M = \pmatrix{\cos(\theta_1/2)  & -\sin(\theta_1/2) \cr
      \sin(\theta_1/2) & \cos(\theta_1/2)}
    \pmatrix{\cosh\lambda & \sinh\lambda \cr
                \sinh\lambda & \cosh\lambda}  \nonumber \\[1ex]
&{}& \hspace{10mm} \times \pmatrix{\cos(\theta_2/2) &
          -\sin(\theta_2/2) \cr \sin(\theta_2/2) & \cos(\theta_2/2)} ,
\end{eqnarray}
with
$$
      \theta_2 = \theta_1 + \alpha_2 .
$$

We can now write this expression as a similarity transformation of a
matrix with equal diagonal elements.  Explicitly,
\begin{equation}
 \pmatrix{\cos(\delta/2)  & -\sin(\delta/2) \cr
    \sin(\delta/2) & \cos(\delta/2)}
 \pmatrix{J & F \cr G & J}
 \pmatrix{\cos(\delta/2)  & \sin(\delta/2) \cr
 -\sin(\delta/2) & \cos(\delta/2)},
\end{equation}
where
\begin{eqnarray}
&{}& \pmatrix{J & F \cr G & J} =
 \pmatrix{\cos(\theta/2)  & -\sin(\theta/2) \cr
    \sin(\theta/2) & \cos(\theta/2)}
 \pmatrix{\cosh\lambda & \sinh\lambda \cr
         \sinh\lambda & \cosh\lambda}        \nonumber \\[1ex]
&{}& \hspace{10mm}\times \pmatrix{\cos(\theta/2)
   & -\sin(\theta/2) \cr  \sin(\theta/2) & \cos(\theta/2)} ,
\end{eqnarray}
with
\begin{equation}
\delta = \frac{1}{2}\left(\theta_1 -\theta_2\right), \qquad
\theta = \frac{1}{2}\left(\theta_1 + \theta_2\right) .
\end{equation}
The above three matrices can be compressed to one equi-diagonal
matrix
\begin{equation}
 \pmatrix{\cosh\lambda\cos\theta &
            -\cosh\lambda\sin\theta + \sinh\lambda \cr
    \cosh\lambda\sin\theta + \sinh\lambda & \cosh\lambda\cos\theta} .
\end{equation}
This matrix can now be written as a similarity transformation of
one of the four Wigner matrices.

When the off-diagonal elements have opposite signs, we can write
this as
\begin{equation}
\pmatrix{e^{\sigma/2} & 0 \cr 0 & e^{-\sigma/2}}
\pmatrix{\cos\omega & -\sin\omega \cr \sin\omega & \cos\omega}
\pmatrix{e^{-\sigma/2} & 0 \cr 0 & e^{\sigma/2}} ,
\end{equation}
with
\begin{eqnarray}
&{}& \cosh\omega = \cosh\lambda \cos\theta ,
     \nonumber \\[1ex]
&{}& e^{2\sigma} = \frac{\cosh\lambda\sin\theta - \sinh\lambda}
                       {\cosh\lambda\sin\theta + \sinh\lambda} .
\end{eqnarray}

When the off-diagonal elements have the same sign, we should write
\begin{equation}
\pmatrix{J & F \cr G & J} = \pmatrix{e^{\sigma/2} & 0 \cr
       0 & e^{-\sigma/2}}
\pmatrix{\cosh\chi & \sinh\chi \cr \sinh\chi & \cosh\chi}
\pmatrix{e^{-\sigma/2} & 0 \cr 0 & e^{\sigma/2}} ,
\end{equation}
with
\begin{eqnarray}
&{}& \cosh\chi = \cosh\lambda \cos\theta ,     \nonumber \\[1ex]
&{}& e^{2\sigma} = \frac{\sinh\lambda - \cosh\lambda\sin\theta}
                   {\cosh\lambda\sin\theta + \sinh\lambda} .
\end{eqnarray}

If one of the off-diagonal elements vanishes, the matrix takes the
form
\begin{equation}
\pmatrix{1 & -2\sinh\lambda \cr 0 & 1} \quad or \qquad
\pmatrix{1 & 0 \cr 2\sinh\lambda & 1} .
\end{equation}

Now the diagonalization of the above expression is straight-forward
according to the mathematical tool we have developed in this paper.
So is the computation of
\begin{equation}
M^N.
\end{equation}
The computation of this process was started in reference~\cite{gk03}.
There, the cycle had to start from the midway in one of the
media, but no explanation was given why.  In this paper, by
introducing the $R(\delta)$ matrix, we can start from an arbitrary
point.  Indeed, this is a physical interpretation of this rotation
matrix.

When the cycle starts from an arbitrary point in the medium, we
can adjust the value of the rotation angle $\delta$ as we did
in Sec.~\ref{cav}.

\section*{Concluding Remarks}

In dealing with a  two-by-two matrix, we are accustomed to
think it can be diagonalized and can be brought to the diagonal
matrix by a rotation.  This paper shows this assumption is not
always true, because the $ABCD$ matrix is not always a rotation
matrix.

It has been shown that the $ABCD$ matrix can be brought to one
of the four one-parameter Wigner matrices through a similarity
transformation, and the similarity transformation is a rotation
followed by a squeeze transformation.

These one-parameter Wigner matrices have the logarithmic property
which allows us to calculate repeated applications by multiplying
the parameter by an integer.   This property is transmitted to
its similarity transformation, since
\begin{equation}
\left(S W S^{-1}\right)^N = S W^N S^{-1} .
\end{equation}

This mathematical technique is applied to laser cavities and
multilayer systems.  However, it is clear from the literature that
the mathematical result stated in this paper is a result of our
efforts to understand physical systems.

This mathematical instrument is applicable to other periodic
systems in physics, such as one-dimensional scattering problems
in quantum mechanics~\cite{sprung93,griff01}, especially in
condensed-matter physics.  We can also broaden our scope to look
into applications in space-time symmetries of elementary particles
in view of the fact that the Wigner matrices used in this paper are
from Wigner's 1939 paper on symmetries in the Lorentz-covariant
world~\cite{wig39,hks86jm}.

For periodic systems, many authors used different approaches.
Sanchez-Soto and his co-authors used the conformal representation
of the $Sp(2)$ group to attack the problem~\cite{monzon00,sanch05}.
This is possible due to the fact that the transformations of this
group can be translated into conformal transformations as noted by
Bargmann~\cite{barg47}.

Another interesting approach to the periodic system is to use
mathematical induction~\cite{kildemo97}.  It is possible to assume
first that the $ABCD$ matrix is known for $N$ cycles, and then
compute the system for $N+1$ cycles.

\end{document}